\documentclass[amsmath,amssymb,aps,byrevtex,superscriptaddress,twocolumn]{revtex4}
\usepackage{epsfig,graphics,graphicx}
\usepackage{color}

\begin{document}
\title{Spontaneous motion of localized structures induced by parity symmetry transition}
\author{A.J. Alvarez-Socorro$^1$, M.G. Clerc} 
\affiliation{Departamento de F\'{i}sica and Millennium Institute for Research in Optics, FCFM, Universidad de Chile, Casilla 487-3, Santiago,
Chile.}
\author{ M. Tlidi} 
\affiliation{D{\'e}partement de Physique, Universit{\'e} Libre de Bruxelles (U.L.B.), CP 231, Campus Plaine, Bruxelles 1050, Belgium.}

\begin{abstract}
We consider a paradigmatic nonvariational scalar Swift-Hohenberg equation that describes  short wavenumber or large wavelength pattern forming systems. 
This work unveils evidence of the transition from stable stationary to moving localized structures in one spatial dimension as a result 
of  a parity breaking instability. This behavior is attributed to the nonvariational character of the model. We show that the nature of this transition is supercritical.
We characterize analytically and numerically this bifurcation scenario from which emerges asymmetric moving localized structures. 
A generalization for two-dimensional settings is discussed.  
\end{abstract}

\maketitle

\section{Introduction} 
Localized structures (LS's) have been theoretically predicted and experimentally observed in many fields of nonlinear science, such as laser physics, 
hydrodynamics, fluidized granular matter, gas discharge system, and biology 
\cite{GlansdorffPrigogine,CrossHohenberg,Nail_lect,Purwins,Clerc,Leblond-Mihalache,Tlidi-PTRA,Lugiatobook,Tlidi_Clerc_sringer,Cates}. 
These solutions correspond to a portion of the pattern  surrounded by regions in the homogeneous steady state. However, localized structures are not necessarily stationary. They can move or exhibit a self-pulsation as a result of
{\it external} symmetry breaking instability induced by  a phase gradient \cite{Turaev_08}, off-axis feedback \cite{Zambrini07},  resonator
detuning \cite{Kestas98}, and space-delayed feedback \cite{Haudin2011}. This motion has been also reported using a selective \cite{Paulau08,Scroggie09} or a regular time-delay feedbacks \cite{Tlidi_99}.

We identify an internal symmetry breaking instability that causes a spontaneous transition from stationary to moving localized structures
 in nonvariational systems. In contrast, variational systems, i.e., dynamical systems characterized by a functional, an internal symmetry breaking instability causes the emergence of motionless asymmetric localized states \cite{Coullet,Burke2007}. 
In order to investigate this nonvariational transition,
 we consider a generic nonvariational scalar  Swift-Hohenberg equation. This is a well-known paradigm in the study of spatial periodic and
localized patterns. It has been derived for that purpose in
liquid crystal light valves with optical feedback \cite{ClercPetrossianResidori2005,Durniak05}, 
in vertical cavity surface emitting lasers \cite{Kozyreff_03}, and in  other fields of nonlinear science \cite{KozyreffTlidi2007}. Generically, it applies to systems that undergo a
Turing-Prigogine instability, close to a second-order critical point marking the onset
of a hysteresis loop. This equation reads
\begin{equation}
\label{Eq-liftschitz_model}
\partial_t u = \eta + \mu u - u^3 -\nu \nabla^{2}u - \nabla^{4}u + 2 b u \nabla^{2} u + c (\nabla u)^2.
\end{equation}
The real order parameter $u=u(x,y,t)$ is an excess field variable measuring the deviation from
criticality. Depending on the context in which Eq.~(\ref{Eq-liftschitz_model}) is derived, the physical meaning of the 
field variable $u$  can be the electric
field, biomass, molecular average orientation, or chemical concentration. The control
parameter $\eta$ measures the input field amplitude, the aridity
parameter, or chemical concentration. The parameter $\mu$ is the cooperativity, and $\nu$ is the diffusion coefficient. 
The Laplace operator $\nabla^{2} \equiv \partial_{xx}+\partial_{yy}$ acts on the  plane $(x,y)$. The parameters 
$b$  and $c$ measure  the strength  of nonvariational effects. 
 The terms proportional to $c$ and $b$, respectively, account for the nonlinear advection and nonlinear diffusion, 
which in optical systems can be generated by the free propagation of feedback light \cite{ClercPetrossianResidori2005,Durniak05}.

\begin{figure*}[t]
\epsfig{figure=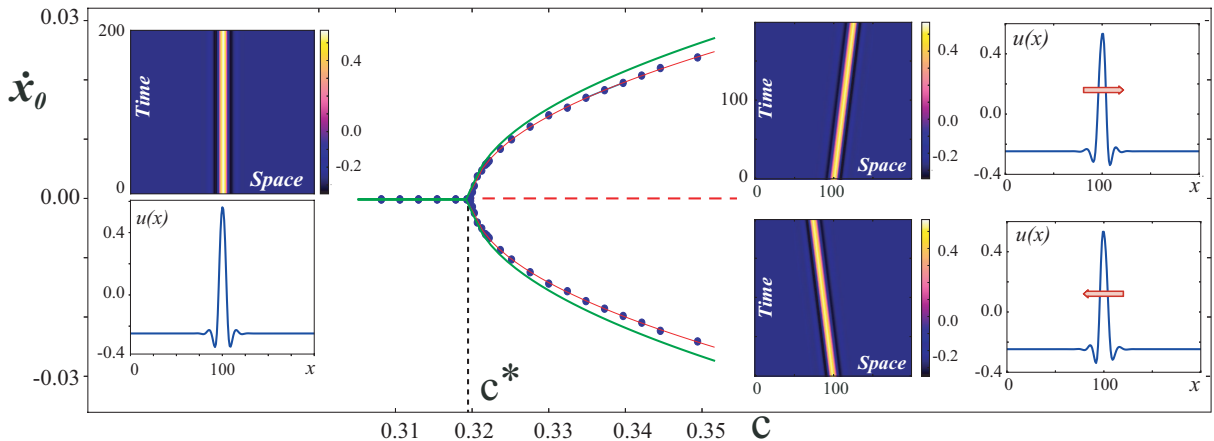,width= 1.85\columnwidth,angle=0}
\caption{(color online) The speed of localized structure $v=\dot{x}_0 $ as a function of the parameter $c$. 
At the transition point $c\equiv c^*\approx 0.319$. Dots indicate localized structures speed obtained from 
numerical simulations of Eq.~(\ref{Eq-liftschitz_model}). Green curves is associated with the analytical results despicted in section III and red curves are associated with the fit of the numerical values given by : 
$v=\dot{x}_0=0.1467\sqrt{c-0.319}$ 
Parameters are  $\mu=\eta=-0.02$, $\nu=1$, and $b=-0.9$. Left insets account for the  profile and the
spatiotemporal evolution of a motionless localized structure. Right top (bottom) insets account for the  profile and the
spatiotemporal evolution of a right (left) moving localized structure.}
\label{Fig-BifurcationDiagram}
\end{figure*}

For $b =  c$, Eq.~(\ref{Eq-liftschitz_model}) is variational \cite{KozyreffTlidi2007}, i.e.,  
the model reads 
$\partial_t u = -\delta F(u)/\delta u,$ with $F(u)$ is the free energy or the Lyapunov functional.
In this case, any perturbation compatible with boundary conditions evolves toward either a 
homogeneous or inhomogeneous (periodic or localized) stationary states corresponding to a local or global minimum of $F(u)$.  
Therefore, complex dynamics such as time oscillations, chaos, and spatiotemporal chaos are not allowed in the limit $b  = c$.  
In particular, in this regime, stationary localized structures and localized patterns have been predicted \cite{Tlidi94}. 
An example of a stationary LS in one-dimension is shown in the left panel of Figure~\ref{Fig-BifurcationDiagram}.
The obtained localized structure has a maximum value of field $u(x,t)$ located at the position $x_0$. 
The stationary LS have been studied for the Eq.~(\ref{Eq-liftschitz_model}) in one dimensional (1D) spatial coordinate, 
as well their snaking bifurcation diagram \cite{Burke2012,Averlant_12}. When $b  \neq c$, the model equation losses its variational structure 
and allows for the mobility of unstable asymmetric localized structures, {\it rung states}, 
that connect the symmetric states \cite{Burke2012}. Indeed, the system exhibits a drift instability 
leading to the motion of localized structures in an arbitrary direction.  However, these rung states are unstable states for small nonvariational 
coefficients.

In this paper, we characterize the transition from stable stationary to moving localized structures in non-variational real Swift-Hohenberg equation.  
 Figure~\ref{Fig-BifurcationDiagram} illustrates stable  moving localized structures.
We show that there exists a threshold over which single LS starts to move in an arbitrary direction since the system is isotropic in both spatial directions. 
We compute  analytically and numerically the bifurcation diagram associated with this transition. In one dimensional setting, 
the transition is always supercritical within the range of the parameters that we explored.   
The threshold and the speed of LS are evaluated both numerically and analytically.  
In two-dimensional settings, numerical simulations of the governing equation indicate that the nature of the transition towards the formation of moving localized bounded states is not a supercritical bifurcation.  It is worth to mention another type of internal mechanism
that occurs in regime devoid of patterns and may lead to a similar phenomenon for fronts propagation through a non-variational Ising Bloch transition \cite{Coullet,Michaelis2001,GilliFrisch94,ClercCoulibalyLaroze2009}. 
The Ising-Bloch transition has been first studied in the context of magnetic walls \cite{Coullet}. Soon after, it has been considered in a various out of equilibrium systems such as driven liquid crystal \cite{GilliFrisch94}, coupled oscillators \cite{ClercCoulibalyLaroze2009},
and nonlinear optic cavity \cite{Taranenko2005}. More recently, it has been shown that non-variational terms can induce propagation of fronts in quasi-one-dimensional liquid crystals based devices  \cite{AlvarezSocorro2017}.
Experimental observation of a supercritical transition from stationary to moving localized structures has been realized in two-dimensional planar gas-discharge systems \cite{Purwins2003}.

The paper is organized as follows, after an introduction, we perform in section II, 
the numerical characterization of the bifurcation scenarios triggered by an internal symmetry breaking instability 
leading to the formation of asymmetric traveling localized structures. At the end of this section, 
we perform numerical simulations in one-dimensional system Eq.~(\ref{Eq-liftschitz_model}). 
In section III, we perform an analytical analysis of the symmetry breaking instability. Two-dimensional moving bounded localized structures are analyzed, and their bifurcation diagram is determined in section IV. Finally, the conclusions are presented  in section V.

\section{Numerical characterization of Parity Breaking Transition }\label{III}

\begin{figure}[t]
\epsfig{figure=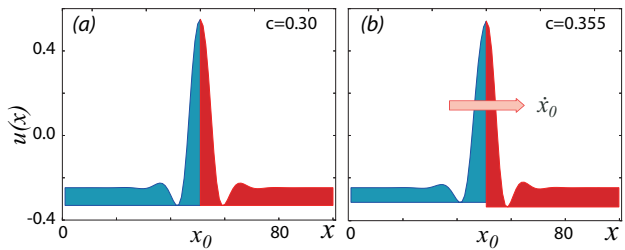,width= 0.95\columnwidth,angle=0}
\caption{(color online) Profile of localized structures obtained from numerical simulations of  Eq.~(\ref{Fig-BifurcationDiagram}). Parameters are $\mu=\eta=-0.02$, $\nu=1$, 
and $b=-0.9$. (a) stationary and  (b) moving  localized structure.}
\label{Assymetric_LS}
\end{figure}

We investigate numerically the model Eq. (1) in the case where $b\neq c$ in 1D with periodic boundary conditions. The results are summarized  in the bifurcation diagram of 
Fig.~\ref{Fig-BifurcationDiagram}.  We fix all parameters and we vary the nonvariational parameter $c$.   When increasing the parameter $c<c^{*}$, LS's are stationary. There exist a threshold $c=c^{*}$ at which transition from stationary to moving LS's takes place. This transition is supercritical. For $c>c^{*}$,  stationary LS becomes unstable, and the system undergoes a bifurcation towards the formation of moving localized structures. The direction in which LS propagates depends on the initial condition used. Indeed, there is no preferred direction since the system is isotropic. The spatial profiles of the stationary and the moving localized states are shown in Fig. \ref{Assymetric_LS}. The shadow regions allows to emphasize the symmetric (stationary) and asymmetric (moving) solutions with respect to the localized structure position. We clearly see from this figure that stationary LS is symmetric with respect to its maximum. This can be explained by the fact that the transition from stationary to moving localized structures is accompanied by a spontaneous spatial parity breaking symmetry. In fact, if the parity with respect to its position $\int_{x_0-L}^{x_0+L} u(x,t)dx$ is positive (negative) 
it moves to the right (left). 

\begin{figure}[t]
\epsfig{figure=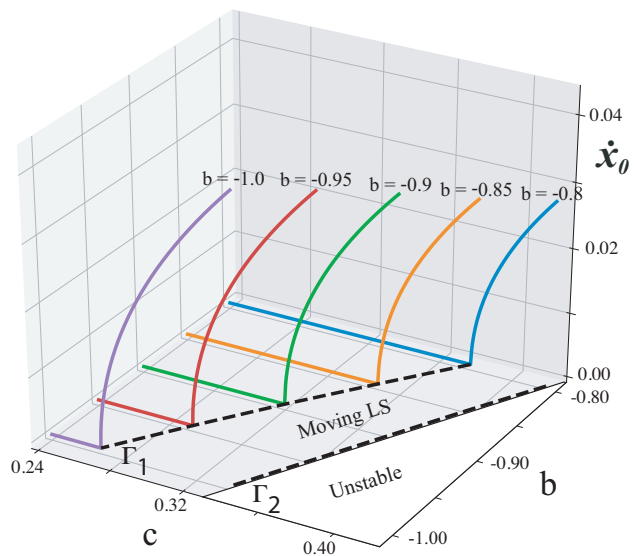,width=0.95\columnwidth,angle=0}
\caption{(color online) The speed   $v=\dot{x}_0 $ as function of the nonvariational parameters $c$ and $b$ 
obtained by numerical simulations of Eq. (\ref{Eq-liftschitz_model}). 
Parameters are $\mu=\eta=-0.02$ and $\nu=1$, for each curve, different fixed value of the 
parameter $b$ are considered, which are indicated in the upper part of the respective curve.
The two segmented curves ($\Gamma_1,\Gamma_2$) limit the region where  moving localized structures are observed.
The line $\Gamma_1$  marks the transition from stationary to moving  localized structures. 
The line indicated the threshold associated with the Andronov-Hopf bifurcation
that destabilizes moving localized structures.}
\label{Fig-Bifurcation_Surface}
\end{figure}

We  have measured numerically the speed of moving LS solutions of Eq.~(\ref{Eq-liftschitz_model}),  by varying both  $c$ and $b$ parameters.  
The results are summarized in Fig.~\ref{Fig-Bifurcation_Surface}. There are three different dynamical regimes. When increasing both parameters 
$b$ and  $c$, the stationary localized structures are stable in the range delimited by the curve  $\Gamma_1$. 
These structures exhibit a spontaneous motion leading the formation of moving LSs solutions in the range 
of parameters delimited by the curves $\Gamma_1$ and $\Gamma_2$. By exploring the parameter space ($b$, $c$), we see clearly from Fig.~\ref{Fig-Bifurcation_Surface} that the bifurcation towards the formation of moving LSs remains supercritical. 

After the numerical characterization, we perform an analytical analysis of the transition towards moving LS. For this purpose, let us consider the linear dynamics  around stationary LS $u_{ls}(x-x_0)$ located at $x=x_0$.  The linear operator reads

\begin{equation}
\begin{aligned}
\mathcal{L}  \varphi \equiv { } & \Big[ \mu - 3u_{ls}^2-\nu \partial_{xx}-\partial_{xxxx}+2b u_{ls}\partial_{xx}\Big] \varphi \\
&  + 2 c (\partial_{x} u_{ls})\partial_{x}\varphi  + 2b (\partial_{xx}u_{ls}) \varphi.
\end{aligned}
\end{equation}

\begin{figure}[t]
\epsfig{figure=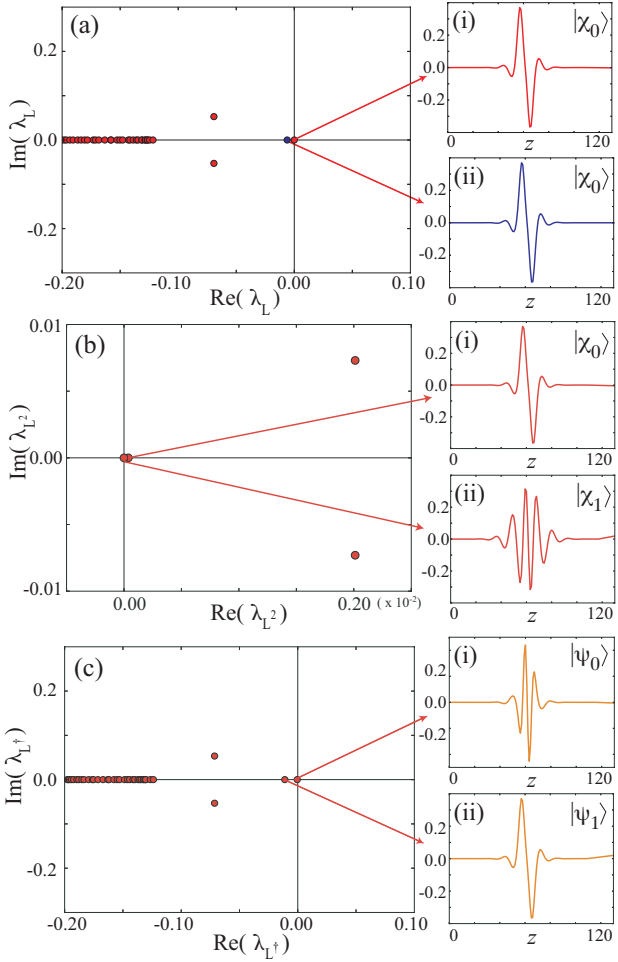,width= 0.95\columnwidth,angle=0}
\caption{(color online) Spectrum of linear operators. Real and imaginary part of the eigenvalues associated with the linear operators (a) $\mathcal{L}$, 
(b) $\mathcal{L}^2$,  and (c) $\mathcal{L}^\dag$. Parameters are  $\mu=-0.02$, $\nu=1.0$, $b=-0.9$ and $c=0.318$. 
Inset figures are the real part of the eigenfunction $|\chi_{0} \rangle$, $|\chi_{1} \rangle$, $|\psi_{0}\rangle$,  and $|\psi_{1}\rangle$ 
for the Goldstone and the asymmetric mode associated with the zero eigenvalue, respectively.}
\label{Fig-Spectra}
\end{figure}

Note that the operator $\mathcal{L}$ is not self-adjoint ($\mathcal{L}\neq \mathcal{L}^{\dagger}$). Due to the lack of analytical solutions of LS for Eq.~(\ref{Eq-liftschitz_model}), we compute numerically the spectrum and eigenvectors associated with 
$\mathcal{L}$, $\mathcal{L}^2$ and $\mathcal{L}^{\dagger}$. The spectrum of ${\mathcal{L}}$ always has an eigenvalue at the origin of the complex plane (the Goldstone mode) as shown in Fig.~\ref{Fig-Spectra}a. The corresponding eigenfunction  denoted by $|\chi_0 \rangle \equiv\partial_{z}u_{ls}(x-x_0)$ is depicted in Fig.~\ref{Fig-Spectra}a (i).  When approaching the parity breaking transition threshold, another mode collides with the Goldstone mode as shown in Fig.~\ref{Fig-Spectra}a. The corresponding eigenfunction of this mode is depicted in Fig.~\ref{Fig-Spectra}a (ii). Note however that the profiles of both eigenfunctions are almost the same. At the threshold these eigenfunctions are identical. This degenerate bifurcation has been reported in the Swift-Hohenberg equation with delayed feedback \cite{Tlidi_99, Sveta-13}.The spectrum of $\mathcal{L}^2$ operator is obtained by using the Jordan matrix decomposition is shown in Fig.~\ref{Fig-Spectra}b. There are two eigenfunctions $|\chi_0 \rangle$ and $|\chi_1 \rangle=u_{as}(x-x_0)$, 
which satisfies 

\begin{equation}
\begin{array}{lll}
\mathcal{L} |\chi_1 \rangle &=& |\chi_0 \rangle, \\
\mathcal{L}^2|\chi_1\rangle &=&0.
\end{array}
\end{equation}

The profiles of  $|\chi_0 \rangle$ and $|\chi_1 \rangle$ are ploted in Fig.~\ref{Fig-Spectra}b (i) and (ii) respectively. From this figure we can see that for  $|\chi_0 \rangle$ mode the integral $\int_{x_0-L}^{x_0+L} |\chi_0 \rangle dx=0$ while for $|\chi_1 \rangle$ the integral $\int_{x_0-L}^{x_0+L} |\chi_1 \rangle dx\neq 0$. This indicate that the profile of $|\chi_1 \rangle$ is assymetric. This assymetric mode has been reported in 
\cite{Coullet,Burke2007, Burke2012,ClercCoulibalyLaroze2009, Purwins2003}.  The eigenvalues associated to the adjoint operator as depicted in Fig.~\ref{Fig-Spectra}c. The eigenfunction associated  $\mathcal{L}^{\dagger}$, $|\psi_0 \rangle$ and  $|\psi_1 \rangle$ are depicted in Fig.~\ref{Fig-Spectra}c (i) and (ii), respectively.

Introducing the canonical inner product

\begin{equation}
\langle g |f \rangle=\int_{-\infty}^\infty  f(x) g(x) dx,
\label{Eq-InnerProduct}
\end{equation}
numerically, we have verified that critical modes are orthogonal $\langle\chi_0 | \chi_1 \rangle=0$.

\section{Analytical description of Parity Breaking Transition}\label{IV}
To provide an analytical understanding of the parity symmetry breaking bifurcation, we focus our analysis on the one-dimensional setting. To do that, we
explore the space-time dynamics in the vicinity of the critical point associated with the transition from stationary to moving LS at $c=c^*$ by defining a small parameter $\epsilon$ which measures the distance from that critical point as  $c=c^*+\epsilon^2  c_0$. Our
objective is to determine a slow time and slow space amplitude
equations.  We expand the variable $u(x,t)$ as
\begin{equation}
\begin{array}{ll}
u(x,t)= & u_{ls}\left(x- x_0(\epsilon t)\right)+\epsilon A'(\epsilon^2 t)u_{as}(x- x_0(\epsilon t))\\
& + w(x, x_0,  A'),
\end{array}\label{Eq-ansatz}
\end{equation}
where $u_{ls}(x- x_0(\epsilon t))$ is the stationary localized structure, 
$x_0$ stands for the position of the localized structure. 
We assume that this position evolves on the slow time scale $\epsilon t$. The function  
$ u_{as}(x- x_0(\epsilon t)) \equiv |\chi_1 \rangle$ is
the generalized eigenfunction corresponding to the asymmetric mode. The amplitude  $A'$ is assumed to evolve on a 
much slower tine scale $\epsilon^2  t$, and $w(x,x_0, A')$ is a small nonlinear correction function 
that follows the scaling $w\ll \epsilon A' \ll 1$. By replacing the above ansatz (\ref{Eq-ansatz})  in the corresponding one-dimensional model 
of Eq.~(\ref{Eq-liftschitz_model})
and linearized in $w$, after straightforward calculations we obtain

\begin{equation}
\begin{array}{lll}
- \mathcal{L} |w \rangle & = &\epsilon \dot{x}_0 |\chi_0 \rangle - \epsilon^3 \dot{ A'} |\chi_1\rangle+\mathcal{L}\epsilon  A' |\chi_1\rangle \\

&&  + c_0 [\partial_{z}u_{ls}|\chi_0\rangle+\epsilon^2  A'^2 \partial_{z}u_{as}  \partial_{z}|\chi_1\rangle  \\

&&+ 2\epsilon  A' \partial_z u_{as}|\chi_0\rangle] -\epsilon^3  A'^3 u_{as}^2 |\chi_1\rangle\\

&& - 3\epsilon^2  A'^2  u_{as} u_{ls}|\chi_1\rangle +c^{*}\epsilon^2  A'^2 \partial_zu_{as} \partial_z |\chi_1\rangle\\

&&+2b(\epsilon A')^2 u_{as}\partial_{zz}|\chi_1\rangle,
\end{array}\label{Eq-LinearW}
\end{equation}
where we have introduced the notation $\dot{x}_0=\partial_t x_0$, $\dot{ A'}=\partial_t  A'$, $z\equiv x-x_0(t)$ 
that  corresponds to the coordinate in the co-moving reference frame 
with speed $\dot{x}_0$, and  $|w \rangle \equiv w(x, x_0,  A',\epsilon)$. 

At order $\epsilon$, the solvability condition \cite{Fredholm} reads
\begin{equation}
\dot{x}_0=- A'.
\label{eq1}
\end{equation}
To determine the equation of the amplitude $A'$ of the asymmetric mode, 
we apply on  Eq.~(\ref{Eq-LinearW}) the linear operator $\mathcal{L}$ and we obtain

$$
\begin{array}{lll}
-\mathcal{L}^2 | w\rangle &=&-\epsilon^3 \dot{A'}\mathcal{L}|\chi_1\rangle+ c_0\mathcal{L} \partial_zu_{ls} |\chi_0\rangle  \\

&&+ c(\epsilon A')^2\mathcal{L} \partial_z u_{as} \partial_z |\chi_1\rangle\\

&& + 2\Delta  c \epsilon   A' \mathcal{L} \partial_z u_{as}  |\chi_0\rangle -(\epsilon  A')^3 \mathcal{L} u_{as}^2|\chi_1\rangle \\

&&+ c^* (\epsilon  A')^2 \mathcal{L} \partial_z u_{as} \partial_z |\chi_1\rangle \\

&&+ 2b(\epsilon  A')^2 \mathcal{L} u_{as} \partial_{zz}|\chi_1\rangle -3 (\epsilon  A')^2 \mathcal{L} u_{as} u_{ls} |\chi_1\rangle.
\end{array}
$$
The application of the solvability condition at the next order leads to
\begin{equation}
\dot{ A'}=\frac{2 c_0 \langle \psi_1|\mathcal{L}\partial_z u_{as} |\chi_0\rangle}
{\langle \psi_1 |\chi_0\rangle} A'
-\frac{\langle \psi_1|\mathcal{L} u_{as}^2 |\chi_1\rangle}{\langle \psi_1 |\chi_0\rangle} A'^3
\label{Pre_Pitchfork}.
\end{equation}
To simplify further equations (\ref{eq1}) and (\ref{Pre_Pitchfork}), we propose the following scaling and change of parameters
\begin{equation}
\begin{array}{l}
\displaystyle A \equiv \frac{\langle\psi_1|\mathcal{L} u_{as}^2 |\chi_1\rangle } {\langle \psi_1 |\chi_0\rangle} A', 
\tau \equiv \frac{\langle \psi_1 |\chi_0\rangle}{\langle\psi_1|\mathcal{L} u_{as}^2 |\chi_1\rangle } t,\\
{ } \\
\displaystyle \sigma \equiv 2  c_0 \frac{\langle \psi_1| \mathcal{L} \partial_z u_{as} |\chi_0\rangle \langle\psi_1| \mathcal{L} u_{as}^2 |\chi_1\rangle} 
{\langle\psi_1|\chi_0\rangle^2}, 
\end{array}
\end{equation}
we get the dynamics for the critical modes
\begin{eqnarray}
\dot{A} (\tau)&=&\sigma A - A^3,\label{Amp:eq1} \\
\dot{x}_0(\tau) &=&  -A. 
\label{Amp:eq2}
\end{eqnarray}
The parameter  $\sigma \propto (c-c^*)$ measures the distance from the critical point associated with the parity symmetry  breaking transition. 
The stationary speed of amplitude Eqs.~(\ref{Amp:eq1},\ref{Amp:eq2}) are $v=\pm \sqrt{\sigma}\propto (c-c^*)^{1/2}$.  This implies that the asymmetric 
mode undergoes at the onset of the instability a drift-pitchfork  bifurcation \cite{Wiggins}, as result of parity breaking symmetry 
\cite{Coullet,Burke2007, Burke2012,ClercCoulibalyLaroze2009, Purwins2003}. 
This bifurcation scenario is in perfect agreement with the results of direct numerical simulations of Eq.~(\ref{Eq-liftschitz_model}) presented in section II [see the bifurcation diagram of Fig.~\ref{Fig-BifurcationDiagram}].

\section{bounded moving localized states in two spatial dimensions}\label{2D}

Most of the experimental observations of localized structures have been realized in two-dimensional systems
\cite{Purwins, Clerc}, in which stationary localized structures are observed.
Experimentally, it has been reported a supercritical transition from stationary to moving localized structures
in a planar gas-discharge system \cite{Purwins2003}.  

\begin{figure}[t]
\epsfig{figure=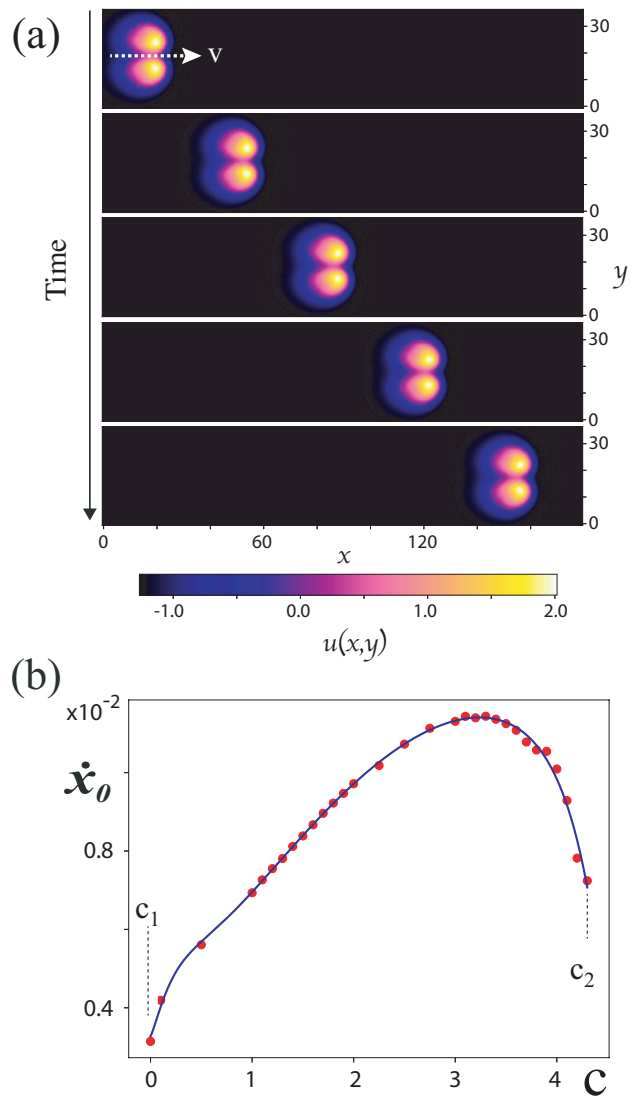,width=0.95\columnwidth,angle=0}
\caption{(color online) Moving bounded localized structures  obtained from numerical simulations of Eq. (\ref{Eq-liftschitz_model}). 
(a)  Temporal sequence  of the moving bounded LSs in $(x,y)$ plane obtained at $t_1=0$, 
$t_2=1500$, $t_3=3000$, $t_4=4500$, $t_5=6000$ time steps. Parameters are $\eta=-2$, $\nu=-1$,  $\mu=-0.092$, 
$b=-2.8$, and $c=3.2$. (b) The speed of   bounded moving
LSs as a function of the parameter $c$. Other parameters are the same as in (a).
The red dots indicate localized structures speed obtained from numerical simulations of Eq.~(\ref{Eq-liftschitz_model}) 
and the blue curve shows an interpolation obtained from these dots.}
\label{Fig-2D_LS}
\end{figure}

\begin{figure*}[t]
\epsfig{figure=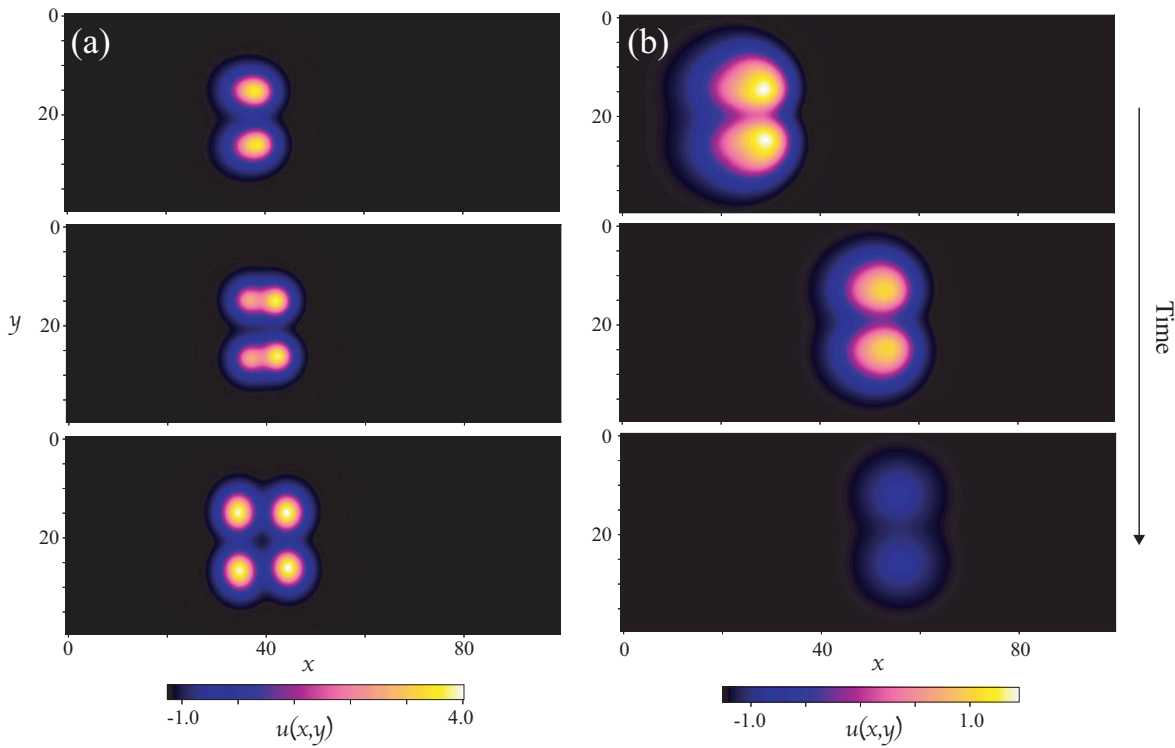,width=1.8\columnwidth,angle=0}
\caption{(color online) Temporal sequence of destabilization of bounded localized structures via (a) a self-replication or (b) a transition towards stationary homogeneous steady states.  Parameters are the same as in Fig. \ref{Fig-2D_LS} except for $c$. (a) $c=-0.2$ and (b) $c=4.5$.}
\label{Fig-2D_LS_UNS}
\end{figure*}

An example of two-dimensional moving localized states obtained by numerical simulations of Eq.~(\ref{Eq-liftschitz_model}) is depicted in Fig. \ref{Fig-2D_LS} (a). In this figure, a time sequence of two-dimensional moving bounded states obtained for periodic boundary conditions is shown. The two spots are bounded together in the course of the motion. The nonvariational effects render this localized moving spots asymmetric. There is no preferent direction for this motion since the system is isotropic in the xy plane. We characterize this motion by computing the speed ($\dot{x_0}$) as a function of the nonvariational parameter $c$. The result is shown in the Fig. \ref{Fig-2D_LS} (b). The existence domain of this moving structures occurs in the range of $c_1 < c < c_2$.  For $c<c_1$ the system undergoes a well documented curvature instability that affect the circular shape of LS and in the course of time leads to self-replication phenomenon. This behaviour is illustrated in Fig. \ref{Fig-2D_LS_UNS} (a). However, for $c>c_2$, bounded localized structures becomes unstable and we observe in this regime transition to homogeneous steady state as shown in Fig. \ref{Fig-2D_LS_UNS} (b).

\section{Conclusion}
\label{V}
We have considered the paradigmatic real nonvariational Swift-Hohenberg equation with cubic nonlinearity. 
We have investigated the transition from  stable stationary to moving localized structures. 
We have shown that the spontaneous motion of localized structures induced by parity symmetry transition and nonvariational effects 
is supercritical and occurs in wide range of  the system parameter values. 
In one dimensional setting, the analytical and the numerical bifurcation diagrams have been established. 
We have derived a normal form equation to describe the amplitude and the speed of moving localized structures. 
We have estimated the threshold as well as the speed of moving asymmetric localized structures. 
A similar scenario has been established 
for cubic-quintic Swift-Hohenberg equation with only 
the nonvariational  nonlinear advective term \cite{Knobloch2011}.
However, in two-dimensional systems, we have shown through numerical simulations that the transition 
towards bounded moving localized state  is rather subcritical. We have shown that there exist a finite range of parameters where bounded LS are stable. Out of this parameter range, the 2D bounded localized state self-replicate  or exhibits transition towards stationary homogeneous steady state. 

Our results are valid in the double limits 
of a critical point associated with nascent bistability and  close to short wavenumber or large wavelength pattern forming  regime.
However, given the universality of model (\ref{Eq-liftschitz_model}), we expect that the transition considered here should be observed in various 
far from equilibrium systems.

\begin{acknowledgments}
M.G.C. and M.T. acknowledge the support of CONICYT project REDES-150046. 
M.T. received support from the  Fonds National de la Recherche Scientifique (Belgium). 
M.G.C. thanks  FONDECYT Projects No. 1150507 and Millennium Institute for Research in Optics. 
A.J.A.-S. thanks financial support from Becas Conicyt 2015, Contract No. 21151618.
\end{acknowledgments}

\end{document}